\documentclass{article} 

\usepackage{arxiv}
\usepackage[utf8]{inputenc} 
\usepackage[T1]{fontenc}    
\usepackage{hyperref}       
\usepackage{url}            
\usepackage{booktabs}       
 \usepackage{amsfonts}       
\usepackage{nicefrac}       
\usepackage{microtype}      
\usepackage{lipsum}
\usepackage[colorinlistoftodos]{todonotes}

\usepackage{afterpage} 
\usepackage{etex}
\usepackage{color}
\usepackage[stable]{footmisc} 
\usepackage[section]{placeins} 
							
\usepackage{graphicx} 
\usepackage{multirow} 
\usepackage{longtable}

\usepackage{amsmath}
\usepackage{amssymb}
\usepackage{amstext}
\usepackage{amsthm}
\usepackage{enumerate}
\usepackage{bbold}
 \usepackage{extarrows}
\usepackage{esvect}

\usepackage{relsize}
\usepackage{array}
\usepackage{dsfont}

\usepackage{pstricks-add}
\usepackage{verbatim} 
\usepackage{caption}
\usepackage{subcaption}	
\usepackage{float}

\usepackage{parskip}

\usepackage{cancel}
\usepackage{framed} 
\usepackage{fancyhdr}
\usepackage{hyperref}

\newcommand{\ket}[1]{\left | #1 \right>}
\newcommand{\R}{\mathbb{R}} 
\newcommand{\Id}{\mathds{1}} 

\newcommand{\br}{\boldsymbol{r}}
\newcommand{\be}{\boldsymbol{e}}

\newcommand{\bn}{\boldsymbol{n}}
\newcommand{\bB}{\boldsymbol{B}}

\newcommand{\bm}{\boldsymbol{m}}

\newcommand{\g}{\gamma} %
\newcommand{\e}{\text{e}} %
\newcommand{\bsigma}{\boldsymbol{\sigma}}

\newcount\colveccount
\newcommand*\colvec[1]{
        \global\colveccount#1
        \begin{pmatrix}
        \colvecnext
}
\def\colvecnext#1{
        #1
        \global\advance\colveccount-1
        \ifnum\colveccount>0
                \\
                \expandafter\colvecnext
        \else
                \end{pmatrix}
        \fi
}

\definecolor{MyYellow}{rgb}{1.0, 0.75, 0.0}
\definecolor{MyGreen}{rgb}{0.0, 0.5, 0.0}
\definecolor{ElectricPurple}{rgb}{0.75, 0.0, 1.0}
\definecolor{CrimsonGlory}{rgb}{0.75, 0.0, 0.2}
\definecolor{DarkRed}{rgb}{0.55, 0.0, 0.0}

\newcommand{\dm}{\boldsymbol{\rho}}
\newcommand{\z}{z}
\newcommand{\x}{x}
\newcommand{\y}{y}
\newcommand{\s}{s}

\begin{document}

\title{Bifurcations and chaos in nonlinear Lindblad equations}
\author{
  Bernd Fernengel  \\
  Institut für Festkörperphysik \\
  Technische Universität Darmstadt \\
  Hochschulst. 6, 64289 Darmstadt, Germany \\
  \texttt{bernd@fkp.tu-darmstadt.de} \\
   \And
  Barbara Drossel  \\
  Institut für Festkörperphysik \\
  Technische Universität Darmstadt \\
  Hochschulst. 6, 64289 Darmstadt, Germany \\
  \texttt{drossel@fkp.tu-darmstadt.de} \\
}
\maketitle
\begin{abstract}
The Lindblad equation describes the dissipative time evolution of a density matrix that characterizes an open quantum system in contact with  its environment. The widespread ensemble interpretation of a density matrix requires its time evolution to be linear. However, when the dynamics of the density matrix is of a quantum system results not only from the interaction with an external environment, but also with other quantum systems of the same type, the ensemble interpretation is inappropriate and  nonlinear dynamics arise naturally. We therefore study the dynamical behavior of nonlinear Lindblad equations using the example of a two-level system. By using techniques developed for classical dynamical systems we show that various types of bifurcations and even chaotic dynamics can occur. As specific examples that display the various types of dynamical behavior, we suggest explicit models based on systems of interacting spins  at finite temperature and exposed to a magnetic field that can change in dependence of the magnetization. Due to the interaction between spins, which is treated at mean-field level, the Hamiltonian as well as the transition rates of the Lindblad equation become dependent on the density matrix.
\end{abstract}

\section{Introduction}
The description of a quantum system by a Schrödinger equation is based on the assumption that the system is sufficiently isolated from the rest of the world that no uncontrollable influences affect its time evolution \cite{Ballentine}. In such a situation, the external environment needs to be taken into account only via boundary conditions, electromagnetic potentials or other potential energy terms. However, when environmental influences that cannot be controlled in detail affect the quantum system, the Schrödinger equation is not appropriate any more \cite{alicki2007quantum,rivas2012open}. As is familiar from statistical mechanics, an environment can induce random transitions between the states of a quantum system so that it does not undergo a unitary time evolution any more. A frequently used equation that captures this effect of the environment is the Lindblad equation \cite{lindblad1976generators}. It is a Master Equation for the density matrix $\dm$ of the quantum system that contains transitions between states in addition to the von Neumann term that describes unitary time evolution.  There are two general ways to derive the Lindblad equation: the first one starts from a quantum description of the system and the bath and takes the trace over the states of the bath in order to obtain a time evolution equation for the reduced density matrix that describes only the system \cite{alicki2007quantum}. This derivation requires a couple of ad-hoc assumptions that cannot really be justified, in particular that the combined state of system and bath can be written as a product state. The second way to derive the Lindblad equation consists in writing down the most general equation that satisfies the requirement that the density matrix remains a density matrix, i.e., that it remains Hermitian, positive semi-definite, and with trace 1. This means that the time evolution must be completely positive and trace preserving \cite{maassen2004quantum}.  Furthermore, in order to preserve the ensemble interpretation of the density matrix, the time evolution is required to be linear in $ \dm$. It can be proven mathematically that the most general equation that satisfies all these criteria is the Lindblad equation.  This second derivation makes no assumption about the nature of the environment, except that its influence on the quantum system depends only on the present state of the quantum system and not on its past, as no memory terms are included. 

Due to the dissipative terms unravelings of Lindblad Equations are non-linear in the wave function and hence they violate superposition principle.
The Lindblad equation is only one of several examples where the theoretical description of a quantum system does not follow unitary time evolution.
In particular in condensed matter theory, methods that violate unitary time evolution are widely employed \cite{drossel2020condensed}. One well-known example are the Hartree and Hartree-Fock theory for many-fermion systems, where the influence of the other particles on a given particle is taken into account via the potential generated by the other particles, which in turn is proportional to their charge density \cite{leggett}. These theories are nonlinear in the wave functions. Applying them to bosons gives the Gross-Pitaevskii equation used for Bose condensates. All these equations are mean-field equations that replace the explicit interaction terms that should occur in a many-particle Schr\"odinger equation by a classical interaction via a potential. 

When such a nonlinear version of the Schrödinger equation is translated into the corresponding von Neumann equation,  the quantum-state dependent terms of the Hamilton operator $H$ cause the commutator $[H,\rho]$ to become nonlinear in the density matrix $\rho$ \cite{breuer2002theory}. This shows that widely used quantum theories can involve time evolutions that are nonlinear not only in the wave function, but also in the density matrix. In fact, Breuer and Petruccione (see Section 3.7 in \cite{breuer2002theory}) list two additional classes of Master equations that are nonlinear in the density matrix: (i) nonlinear Boltzmann equations, which describe the time evolution of a one-particle density matrix due to collisions with other particles, (ii) mean-field Master Equations for $n$ identical interacting quantum systems, where the average influence of the other systems on the time evolution of the density matrix of one system is included in a similar way as in the Hartree method. 

A discrete version of nonlinear dynamics of the density matrix is obtained via carefully designed quantum state transformations, as for instance described by Bechmann et al.~\cite{bechmann1998non}. In such transformations, several spins are prepared in specified states and are then subjected to an interaction that causes them to become entangled (for example via a C-NOT gate). When then a measurement with an additional post-selection is performed, the resulting quantum state transformation is nonlinear. By applying such a quantum state transformation iteratively, one can obtain complex dynamics and even chaos \cite{kiss2006complex, kiss2011measurement, torres2017measurement}. By chaos, the authors mean exponential sensitivity of the dynamics of the density matrix with respect to the initial condition. 

By comparing this quantum state transformation system that shows a discrete dynamics of the density matrix with the above-mentioned three types of systems that show a continuous nonlinear dynamics of the density matrix, we can identify two shared features: The first feature is an interaction  between identical particles that are part of the quantum system of interest. Since the density matrix for all these models is that of one quantum particle, this interaction makes the interpretation of the density matrix as an ensemble of independent systems invalid. The  second feature is the interaction with a classical external world that destroys unitary time evolution. In the simplest cases, the role of this external world consists merely in imposing a temperature on the system and causing it to relax to thermal equilibrium (for instance for systems described by a Boltzmann equation). In the quantum information example, this external world consists in a device for state preparation and measurement/postselection and thus establishes a nonequilibrium situation, which in turn is a prerequisite for obtaining dynamical trajectories that do not relax towards fixed points. 

Another way of establishing a nonequilibrium situation consists in periodic driving. Indeed, a periodically driven quantum many-particle system was found to show periodic attractors, bifurcations, and chaos when a mean-field calculation of the dynamics was performed  \cite{hartmann2017asymptotic}. 

This widespread occurrence of time evolutions that are nonlinear in the density matrix suggests that bifurcations and nontrivial attractors can occur generically in open quantum systems if they are driven away from equilibrium, although the quantum state transformation example suggests that this requires a careful design of the system. In this paper, we therefore want to explore in a more general way the occurrence of various types of bifurcations, and in particular of limit cycles and chaos within the framework of the Lindblad equation. We will focus on the simplest possible quantum system, namely a two-level system, or, equivalently, a spin-1/2. We will present two different types of approaches: The first one will be a generic demonstration that various well-known types of bifurcations can be obtained by making the transition rates in the Lindblad equation dependent on the density matrix and ensuring that the density matrix remains positive semi-definite with trace one under the resulting time evolution. This includes the occurrence of limit cycles and chaos. The second one will be a more microscopic approach, where a series of explicit physical model will be constructed in order to illustrate that one can indeed conceive of real-world systems that show these nonlinear dynamics phenomena. The ingredients of these explicit models are the following: First, we include an interaction between the individual quantum systems that constitute the ensemble, having mostly the spins of a solid from condensed matter physics in mind. The interaction between these spins, which will be taken into account on the level of a mean-field approximation, leads to nonlinearities, making the energy eigenstates of the spins dependent on the density matrix. By applying a finite temperature, the dynamics of the Lindblad equation consists in spin flips between and projections onto energy eigenstates of the spins.  In order to create an ongoing nonequilibrium situation, we couple the spins to an environment that measures the magnetization and responds by applying a magnetic field the strength and direction of which depend on the measured magnetization, which in turn is described by the density matrix.

\section{The Lindblad equation}
The Lindblad equation describes the time evolution of the density matrix $ \dm = \dm(t)$ of an open quantum system in the limit where no memory terms are required, \cite{breuer2002theory} 
\begin{equation} \label{eq_Lindblad_rho_0}
\partial_t \dm = \underbrace{-\frac{i}{\hbar} [H, \dm]}_{\text{von Neumann term}} + \underbrace{\sum_k \gamma_k \left( A_k \,  \dm \, A_k^\dagger - \frac{1}{2} \{A_k^\dagger \,  A_k, \dm \} \right)}_{\text{dissipator}}\, ,
\end{equation}
with $H$ being the Hamilton operator, $[\cdot, \cdot]$ and $\{\cdot, \cdot\}$ denoting the commutator and  anticommutator respectively, and the $A_k$ being the Lindblad operators that introduce dissipation into the quantum system. They can be chosen such that they form an orthonormal basis, that is $\text{trace}(A_k^\dagger A_j)=\delta_{kj}$. Below, we will only consider examples where the $A_k$ are trace-free. This is no real restriction as a model with nonvanishing traces of the Lindblad operator can be mapped onto one with vanishing traces by modifying the Hamilton operator \cite{breuer2002theory}.
We will consider a two-state model. This can represent any two-level system, but the examples discussed below are best suited for a spin-1/2 system. 
 In the following, we will use the non-diagonal form of the Lindblad equation that is obtained by expressing the Lindblad operators in a given basis of trace-free matrices,
\begin{align}
L_1 :=  \begin{pmatrix}
0 & 0\\
1 & 0
\end{pmatrix}, \;
L_2 := \begin{pmatrix}
0 & 1\\
0 & 0
\end{pmatrix}, \; 
L_3 := \begin{pmatrix}
1 & 0\\
0 & -1
\end{pmatrix}.  \label{eq_L1-3}
\end{align}
By writing $A_k = \sum\limits_{i=1}^3 a_i^{(k)} L_i$, the Lindblad equation becomes
\begin{align} \label{eq_Lindblad_rho}
\dot{\dm} =-\frac{i}{\hbar}[H, \dm] + \, \sum\limits_{i,j=1}^3 h_{i j}\left( L_i \, \dm\, L_j^\dagger - \frac{1}{2} \{L_j^\dagger L_i, \dm \} \right),  
\end{align}
with $h_{ij} = \gamma_k \, a_i^{(k)} a_j^{(k)^* }= h_{ji}^*$  being positive semi-definite. 

In the standard Lindblad equation, the transition rates $\gamma_k$ and the Lindblad operators $A_k$ are fixed model ingredients that capture the effect of the environment on the quantum system. We will, however, consider the possibility that the transition rates and Lindblad operators depend on the density matrix $\dm$, as explained briefly in the Introduction, and as made explicit by the specific models given further below. In the representation \eqref{eq_Lindblad_rho}, the dependence of the $A_k$ on the density matrix goes into the coefficients $a_i^{(k)}$, as does the dependence of the $\gamma_k$ on the density matrix. 

It is useful to write $\dm$ in terms of three parameters,
\begin{align} \label{rhoxyz}
\dm = \frac{1}{2} \, \left( \Id + \boldsymbol{x} \cdot \boldsymbol{\sigma} \right) =  \frac{1}{2} \begin{pmatrix}
1 + z         & x - i \, y \\
x + i \, y & 1 - z
\end{pmatrix}, 
\end{align}
where $\boldsymbol{\sigma}$ is the vector of Pauli spin matrices, and $\x$, $\y$ and $\z$ are real-valued. The only constraint on their values is that $(\x,\y,\z)$ must stay inside the admissible region $U $,
which assures that $\dm$ stays positive semi-definite,
\begin{align}\label{AdmissibleRegion}
U := \left\{\, (\x,\y,\z) \in \R^3 \, , |\, \x^2 + \y^2 + \z^2\, |\leq 1 \,\right\}.
\end{align} 
By defining  
$\Gamma := \frac{1}{2}\left( h_{11} + h_{22}  + 4 \, h_{33} \right) $, we can write equation \eqref{eq_Lindblad_rho} in the following form,
\begin{eqnarray} \label{eq_Lindblad_components}
\dot{\z} &=& \left( h_{11}-h_{22}\right) - \left(h_{11} + h_{22}\right) \,\z + \text{Re}[h_{23} + h_{13}] \, \x + \text{Im}[h_{23} - h_{13}] \, \y + \frac{-2 \, \text{Im} [ H_{10}]\,\x + \, 2 \, \text{Re} [H_{10}]\, \y}{\hbar}  , \\ 
\dot{\x} &=&  2 \, \text{Re}[h_{23} - h_{13}] + \left(\text{Re}[h_{23} + h_{13}]\right)\, \z + \left(\text{Re}[h_{12}]-\Gamma \right)\, \x - \left(\text{Im}[h_{12}]\right) \,\y + \frac{ 2 \, \text{Im} [H_{10}]\, \z - (H_{00}-H_{11})\, \y}{ \hbar}\, , \nonumber \\ 
\dot{\y} &=& 2 \,  \text{Im}[h_{23} + h_{13}] + \left(\text{Im}[h_{23} - h_{13}]\right) \, \z - \left(\text{Im}[h_{12}]\right)\, \x - \left(\text{Re}[h_{12}] + \Gamma\right) \,\y +  \frac{ -2 \, \text{Re} [H_{10}]\,\z + (H_{00}-H_{11})\, \x\,}{\hbar} .\nonumber
\end{eqnarray}
The last terms are due to the von Neumann term, with $H_{ij}$ being the matrix elements of the Hamiltonian with respect to the two basis states $\ket{0}$ and $\ket{1}$.
As we assume that the transition rates and the Hamilton operator can depend on the density matrix, we have in general $h_{ij}  = h_{ij}(\x, \y, \z)$ and $H=H(\x, \y, \z)$. We can ensure that the matrix $h_{ij}$ remains positive semi-definite at all times either by calculating explicitly the relevant constraints on the parameters of the model, or by using the diagonal form \eqref{eq_Lindblad_rho_0} with positive transition rates $\gamma$.

 Below,  we will consider only special cases where part of the terms vanish or have a simpler form. We will start with the case that the system of  equations \eqref{eq_Lindblad_components} is one-dimensional, where we can already investigate pitchfork bifurcations and saddle-node bifurcations.
 When we go to the two-dimensional case, we will additionally obtain Hopf bifurcations and limit cycles, and the three-dimensional case will furthermore yield strange attractors.

\section{One-dimensional case}
 \label{SubSection_1_dim_case}
The system of equations (\ref{eq_Lindblad_components}) becomes one-dimensional when the transition matrices $A_k$  make transitions between and measurements of the two eigenstates of the Hamilton operator. If we choose as basis states the two eigenstates of the Hamilton operator, the matrices $A_k$ become identical to the operators $L_1$ to $L_3$, and the coefficient matrix $h$ becomes diagonal. Furthermore, a diagonal density matrix $\dm$ remains diagonal under time evolution, and the von Neumann term vanishes. 
In this situation, equations \eqref{eq_Lindblad_components} reduce to the equation
\begin{equation}\label{eq_Lindblad_components_2}
    \dot{\s} =  -  \Gamma \, \s \, 
\end{equation}
for the of{f-}diagonal elements  $\s = \x + i \,  \y$ and
\begin{equation}
\label{Lindblad_1dim_z}
\dot z = \left(h_{11}-h_{22}\right) -  \left( h_{11} + h_{22}\right) \, \z \, ,
\end{equation}
which determines the time evolution of the diagonal elements.
As the off-diagonal entries of the density matrix decrease to zero, it is sufficient to investigate the time evolution of $z$. 

\begin{figure}[H]
\begin{center}
\begin{subfigure}{0.39\textwidth}
   \subcaption{}
      \label{Fig_1_a}
  \includegraphics[width=0.8\columnwidth]{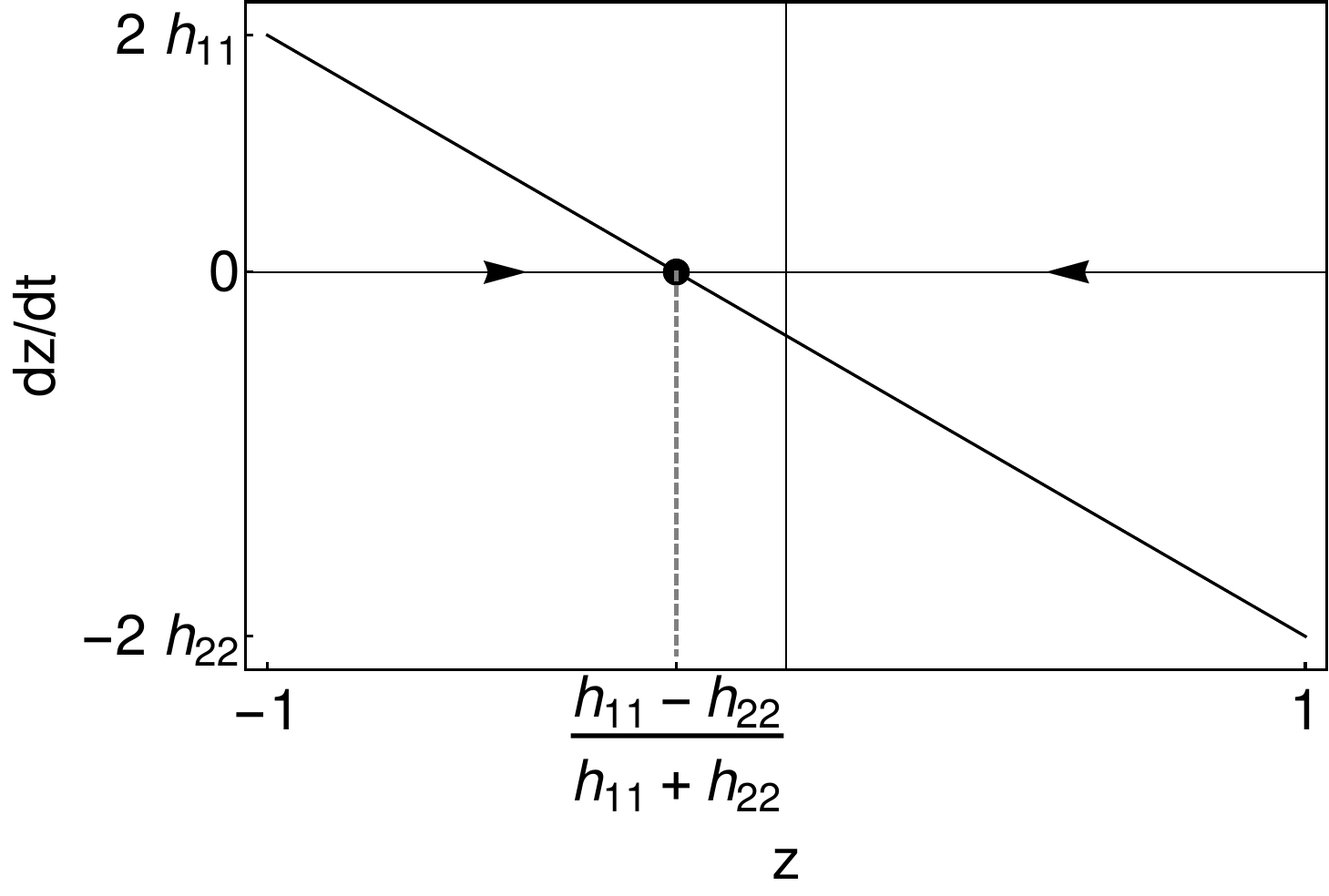} 
\end{subfigure}
\begin{subfigure}{0.39\textwidth}
   \subcaption{}
   \label{Fig_1_b}
  \includegraphics[width=0.8\columnwidth]{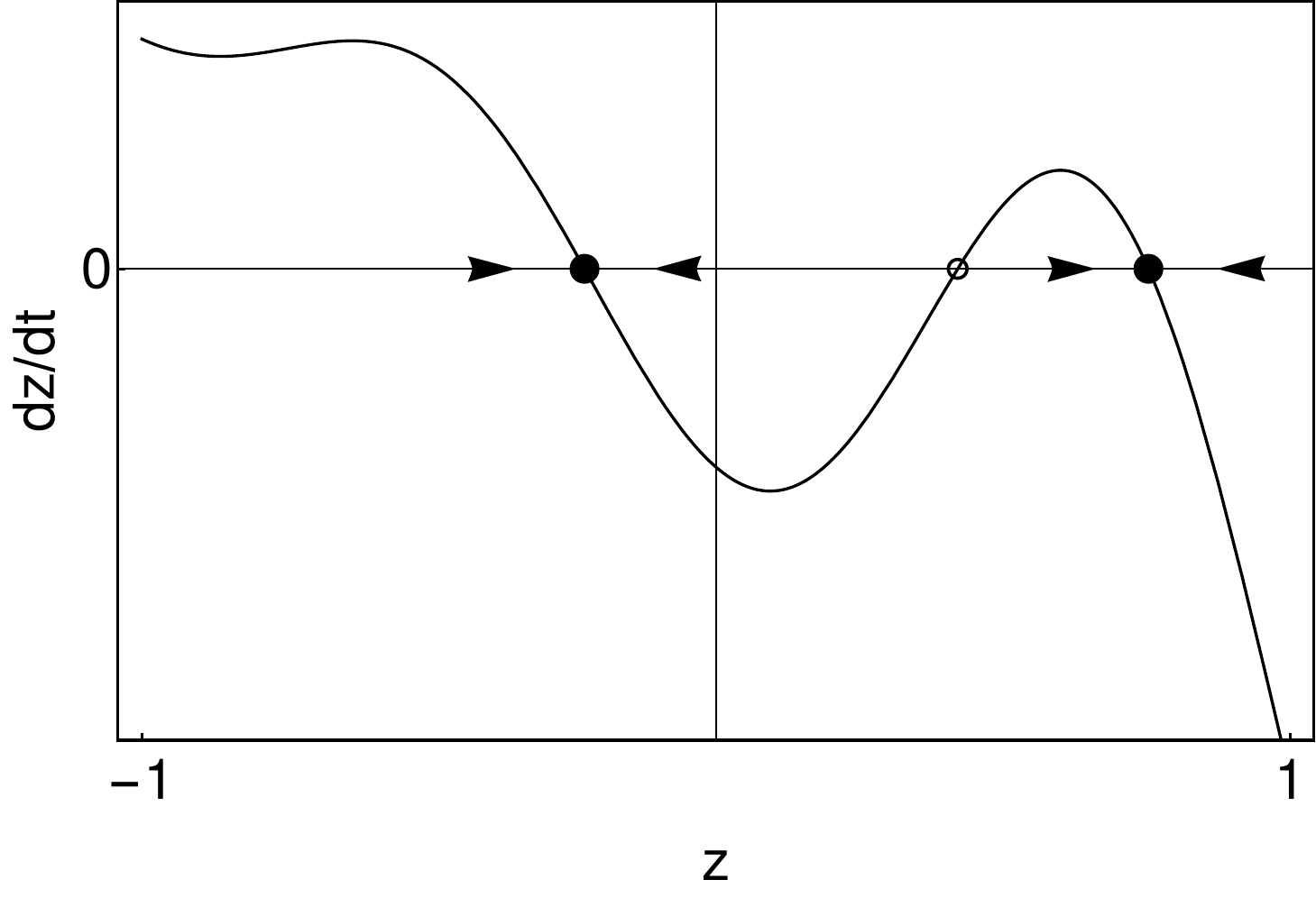} 
\end{subfigure}
\caption{Example phase portrait of the dynamical system described by equation (\ref{Lindblad_1dim_z}). (a) When $\dot z$ is linear in $z$, only one fixed point occurs. (b) When $\dot z$ becomes nonlinear in $z$, the system can have several fixed points.  Stable (unstable) fixed points are marked by dots $\bullet$ and circles $\circ$, respectively. }
\label{Fig_1_PhasePortrait_WithAndWithoutFeedback} 
\end{center}
\end{figure}

When the transition rates are independent of $\dm$, the time evolution goes to
\begin{equation}
\lim\limits_{t\to\infty} \dm(t) =  \,\frac{1}{h_{11} + h_{22}}
\begin{pmatrix}
h_{11} & 0 \\
0      & h_{22}
\end{pmatrix} \, . 
 \end{equation}
If the transition rates depend on $z$, several fixed points can occur, as illustrated in Figure \ref{Fig_1_PhasePortrait_WithAndWithoutFeedback}(b), and a parameter change can lead to various bifurcations as fixed points are created or destroyed. 

In the following, we will perform a generic study of the pitchfork bifurcation and the saddle-node bifurcation, and we will build specific physical systems that will display these bifurcations. 


\subsection {Pitchfork bifurcation }
When the system has an intrinsic symmetry around $\z = 0$,  equation (\ref{Lindblad_1dim_z}) contains only odd powers when written as a polynomial of $\z$. Such a symmetry occurs for instance for a two-level spin system when none of the two spin orientations is preferred by the environment. The minimum model showing this pitchfork bifurcation is given by a polynomial of the order 3. Through an appropriate choice of the time scale, this polynomial can be brought to the form 
\begin{align} \label{eq_pitchfork}
\dot{\z} = - \z \cdot \left( t + \z^2 \right),
\end{align}
which is the normal form of the pitchfork bifurcation and depends only on the parameter $t$, with a stable fixed point at $z=0$ for $t>0$ and two stable fixed points at $z=\pm \sqrt {-t}$ for $t<0$.
An explicit form for the transition rates is in this case
\begin{align} \label{Ansatz_hii_Pitchfork}
h_{11} (\z) =& \, \alpha + \left( \alpha - \frac{t}{2} \right) \, \z + \frac{1}{2} \z^2 \\
h_{22} (\z) =& \, \alpha - \left( \alpha - \frac{t}{2} \right) \, \z + \frac{1}{2} \z^2. \nonumber
\end{align}
This ansatz respects the symmetry between the states $\ket{0}$ and $\ket{1}$ and their transition rates, namely $h_{11}(\z) = h_{22}(-\z)$. 
A sufficient criterion for the transition rates  $h_{ii}(\z)$ being non-negative is that the parameters $t$ and $\alpha$ satisfy $\alpha \in (0,2)$, $ \vert t - 2 \, \alpha \vert \leq \sqrt{8 \, \alpha}$.  
The normal form of the pitchfork bifurcation (expression \eqref{eq_pitchfork}) can only be a good approximation in a sufficiently small region around this bifurcation. 
We will see this in the following example.

An explicit, microscopic model is given by an ensemble of spins that are associated with a magnetic moment,  which are embedded in a medium, for instance a crystal. They are coupled to each other and experience flips due to the interaction with the phonon heat bath of temperature $T$ of the medium. We interpret now the density matrix $\dm { \,=\, \frac{1}{2} \left(\Id + \br\cdot\bsigma\right) }$ as describing the ensemble of these spins, assuming that their number is so large that stochastic fluctuations of the state of this ensemble can be neglected. The vector ${ \br\,=\, }(x,y,z)$ is in this case proportional to the magnetization { $\bm$} of the spin system, and simultaneously it is the averaged magnetic moment of a single spin that flips between its two orientations. The Hamilton operator reads
\begin{equation}\label{Hmagnetic}
H = - \frac{1}{2} \sum\limits_{ij}J_{ij} \,  \bsigma_i \cdot \bsigma_j = 
-\frac{1}{2} \sum_i \bsigma_i \cdot\sum_j J_{ij} \, \bsigma_j
\approx -\frac{1}{2} \sum_i \bsigma_i \cdot \left\langle\sum_j J_{ij}  \; \bsigma_j\right\rangle_i =
- \frac{1}{2} \sum_i\alpha\, \bm  \cdot \bsigma_i \equiv \frac{1}{2} \sum_i \hat h_i ,
\end{equation}
where we have made the mean-field approximation that each spin sees the same local field $\alpha\,  \bm$ caused by the other spins. We choose the normalization of $\bm$ such that its components are identical to the entries of the density matrix representing the ensemble of spins, i.e., $\bm = (x,y,z)$. 

An effectively one-dimensional model is obtained if the embedding medium imposes an easy axis on the system, as is the case in uniaxial magnetic systems. Orientation of the spins along this axis is energetically favored in such systems. We implement in the following the easy axis by the choice of Lindblad operators. By choosing $L_1$ to $L_3$, see \eqref{eq_L1-3}, the easy axis is the $z$ axis. We  have already shown above that in this case the off-diagonal elements of the density matrix relax to zero (see \eqref{eq_Lindblad_components_2}), and that the dynamics of the diagonal elements, i.e., the dynamics of $z$, decouples from the other two equations and is described by \eqref{Lindblad_1dim_z}. We consider therefore in the following only the $z$ dynamics. 

By setting $\bm = (0,0,z)$, the two eigenvalues of the single-spin hamiltonian $\hat h$ are  $\pm\, \alpha z $.  
The interaction with the heat bath causes the spins to flip between these two energy eigenstates, which means that the Lindblad equation contains the two Lindblad operators $L_1$ and $L_2$. The operator $L_1$ makes a spin flip from the spin orientation in positive $z$ direction to that in negative $z$ direction, while $L_2$ makes the reverse flip. 
The flip rates must be chosen such that we obtain Boltzmann weights in thermal equilibrium, leading to
\begin{equation} 
h_{11} =\gamma_1= \gamma\frac{e^{-\beta\alpha z}}{e^{-\beta\alpha z} + e^{\beta\alpha z}} \quad \text{ and } \quad  h_{22} =\gamma_2 = \gamma\frac{e^{\beta\alpha z}}{e^{-\beta\alpha z} + e^{\beta\alpha z}}
\label{eq_fliprates}
\end{equation}

with $\beta = 1/k_BT$. Since the spins flip between their energy eigenstates, the von Neumann term,  which is obtained by using the single-spin hamiltonian (see equation \eqref{Hmagnetic}) $\hat h = -\alpha\bsigma \cdot \br = -\alpha z \sigma_z$, vanishes.

The dynamical equation \eqref{Lindblad_1dim_z} for $z$ then becomes
\begin{equation}\label{dotzpitchfork}
    \dot z = \gamma  \tanh (\beta\alpha z) - \gamma  z\, ,
\end{equation}
For $\alpha \, \beta \leq 1$ , equation \eqref{dotzpitchfork} has only one fixed point at $z=0$, which is stable. For $\alpha \, \beta > 1$, the fixed point at $z=0$ is unstable, and there are two additional fixed points $\pm\, z^*$ that are given by $\tanh(\alpha\,\beta\,z^*) = z^*$. These fixed points are stable. 
 We thus have a pitchfork bifurcation at $\beta\alpha = 1$. 

Not surprisingly, our calculation has resulted in a  phase transition to ferromagnetism, as it has used the ideas behind the mean-field theory by Weiss. When $\beta\alpha$ is close to 1, $z$ is small, we have  $\tanh(x) = x - x^3/3 + \mathcal{O}(x^5)$, and the normal form \eqref{eq_pitchfork} of the bifurcation is a good approximation, but for larger values of $z$ higher-order terms of the Taylor expansion become important.

\begin{figure}
\begin{center}
\begin{subfigure}{0.49\textwidth}
  \subcaption{  }
\label{PitchFork_Diagram_LessThan}
   \hspace*{9mm} \includegraphics[width=1.0\columnwidth]{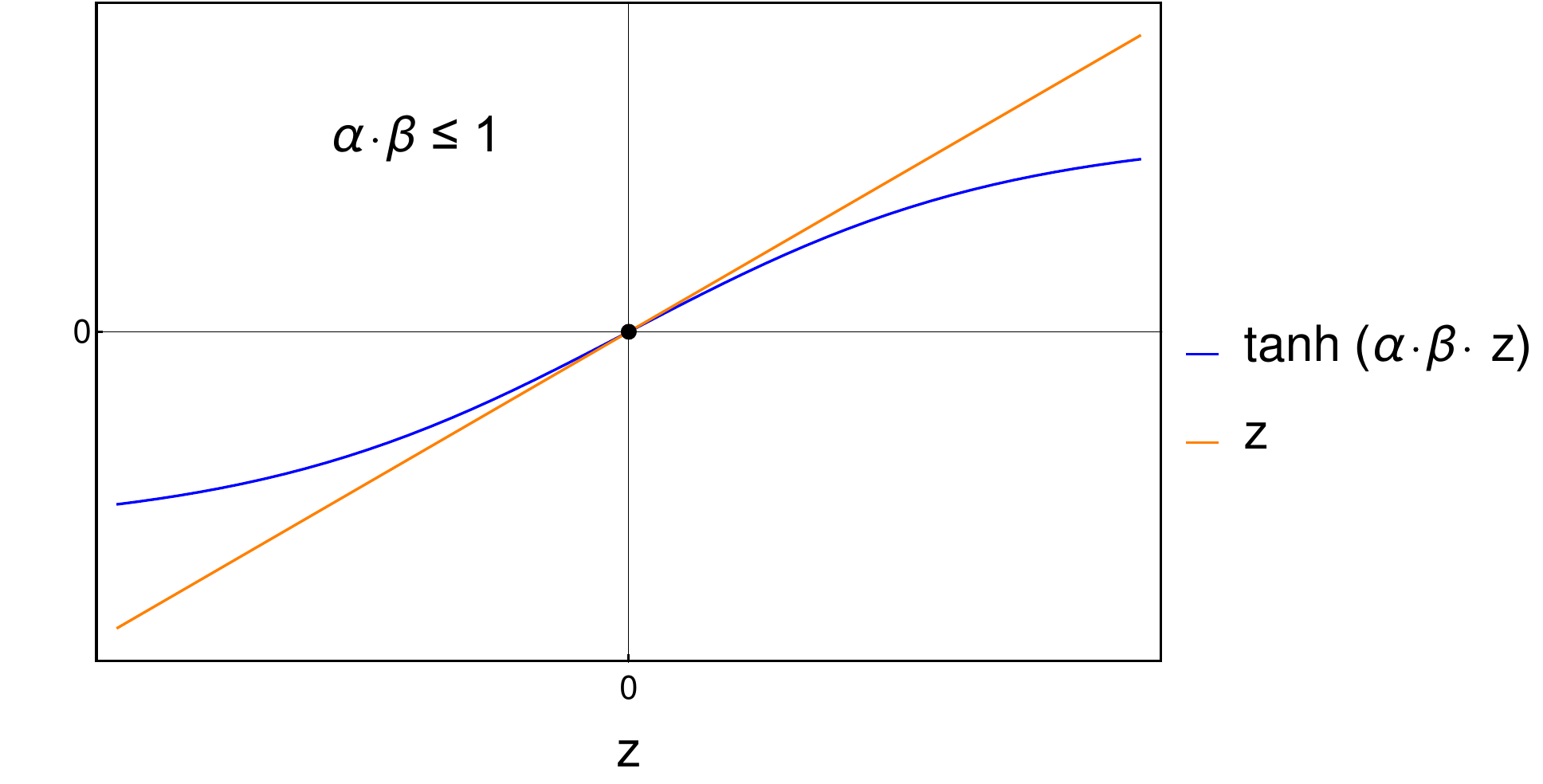} 
\end{subfigure}\begin{subfigure}{0.49\textwidth}
  \subcaption{}
  \label{PitchFork_Diagram_LargerThan}
   \hspace*{10mm} \includegraphics[width=0.75\columnwidth]{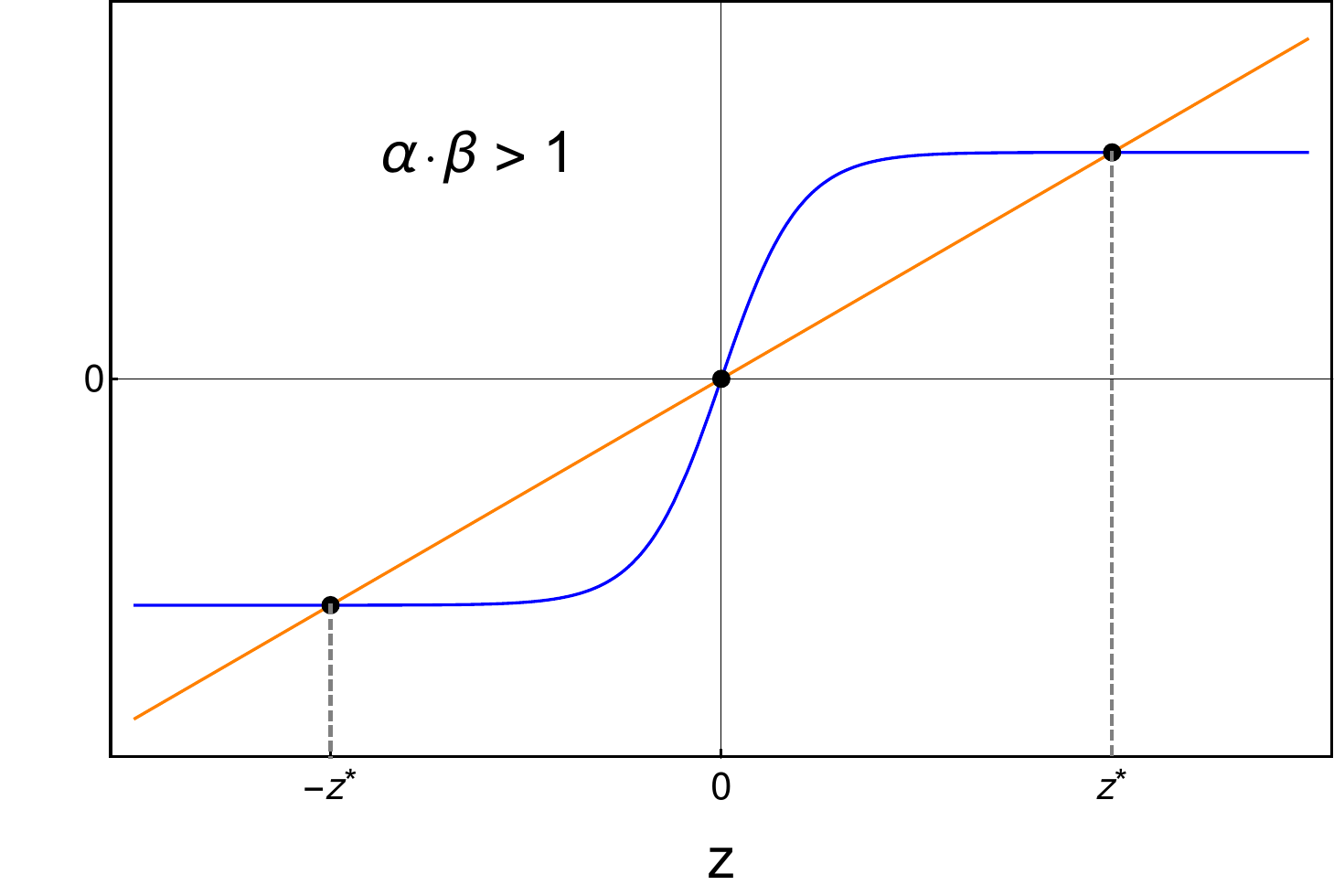} 
\end{subfigure}
\caption{Graphical illustration of the right hand-side of equation \eqref{dotzpitchfork}: The dynamical fixed points are the fixed points of the function $z\mapsto \tanh(\alpha\,\beta\,z)$. For $\alpha \, \beta \leq 1$, (Figure \ref{PitchFork_Diagram_LessThan}) $z=0$ is the only fixed point, which is stable. For $\alpha \, \beta > 1$ (Figure \ref{PitchFork_Diagram_LargerThan}) the fixed point at the origin is unstable and there are two additional, stable fixed points $\pm\,z^*$.  }
\label{PitchFork_Diagram} 
\end{center}
\end{figure}

\subsection { Saddle-node bifurcation }
When the system contains no symmetry around $\z=0$, we can add a constant term to the right-hand side of equation (\ref{eq_pitchfork}), resulting in 
\begin{align} \label{eq_SaddleNode}
\dot{\z} = - \z \cdot \left( t + \z^2 \right) + b . 
\end{align}
With the choice $t>0$, this system has 3 fixed points for $b=0$.
When the parameter $b$ crosses a critical value $b_c = \pm 2 \left( \frac{|t|}{3} \right)^\frac{3}{2}b_c$, a saddle-node bifurcation happens as the unstable fixed point collides with one of the stable fixed points, with the outcome that both of them vanish.

\begin{figure}[H]
\begin{center}
\includegraphics[width=0.6\columnwidth]{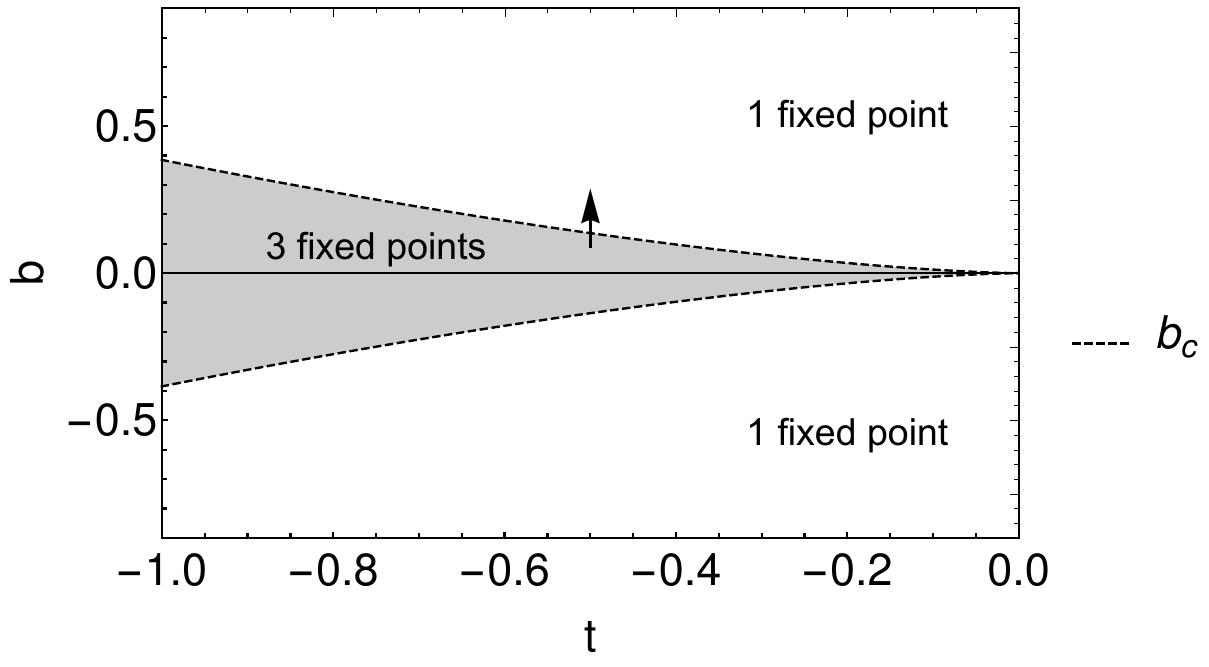} 
\caption{Stability diagram for model \eqref{Ansatz_hii_SaddleNode}. The dashed line indicates the critical value $ b_c = \pm 2 \left( \frac{|t|}{3} \right)^\frac{3}{2}b_c$. We have three fixed point (two of which are stable) in the gray area and one fixed points in the white areas. The saddle node bifurcation occurs when crossing the dashed line, as indicated by the arrow.  }
\label{SaddleNode_Parameter_Diagram}
\end{center}
\end{figure}

The bifurcation equation (\ref{eq_SaddleNode}) can be expressed in the form of  equation (\ref{Lindblad_1dim_z}) with non-negative functions $h_{ii} (q)$ for instance by setting
\begin{align} \label{Ansatz_hii_SaddleNode}
h_{11} (\z) =& \, \alpha + \frac{b}{2} + \left( \alpha - \frac{t}{2} \right) \, \z + \frac{1}{2} \z^2 \\
h_{22} (\z) =& \, \alpha - \frac{b}{2} - \left( \alpha - \frac{t}{2} \right) \, \z + \frac{1}{2} \z^2. \nonumber
\end{align}
The parameters $\alpha$, $t$ and $b$ must satisfy
\begin{align} \label{SaddleNode_ParameterRegions}
\alpha &\in \,(0,2),  \\
\vert t - 2 \, \alpha \vert &\leq \, \sqrt{8 \, \alpha} \nonumber \\
\vert b \vert &\leq\, 2 \, \alpha. \nonumber 
\end{align}
Figure \ref{SaddleNode_Parameter_Diagram} shows the stability diagram of this system. 

An explicit microscopic model that shows such a saddle-node bifurcation is obtained when model \eqref{Hmagnetic} is supplemented by an external magnetic field $\bB$, resulting in the single-spin hamiltonian
\begin{equation}
    \hat{h} = - \alpha \bm  \cdot \bsigma - \mu \bB \cdot \bsigma\, .\label{eq_hmb}
\end{equation}
We choose $\bB=(0,0,B)$ so that it points along the easy axis in positive $z$ direction. As before, we implement the easy axis in the simplest possible way by choosing $L_1$ and $L_2$, see \eqref{eq_L1-3} as Lindblad operators. Due to the von Neumann term, see equation \eqref{eq_Lindblad_components},  \eqref{eq_Lindblad_components_2} is now extended to
\begin{equation}
    \dot s = -\Gamma s -i\frac{2\mu}{\hbar}B s\, ,
\end{equation}
which relaxes again to $s=0$. The $z$ dynamics, given by \eqref{Lindblad_1dim_z}, now takes the form 

\begin{equation}\label{zdotsaddlenode}
    \dot z =  \gamma  \tanh (\beta(\alpha z + \mu B)) - \gamma  z\, ,
\end{equation}
which shows saddle-node bifurcations as $B$ or $T$ are changed. 
For parameter values such that $z$ is small at the fixed points, we  can again set $\tanh( x) =  x - x^3/3 + \mathcal{O}(x^5)$ and regain the form \eqref{eq_SaddleNode}.

\begin{figure}[H]
\begin{center}
  \includegraphics[width=0.7\columnwidth]{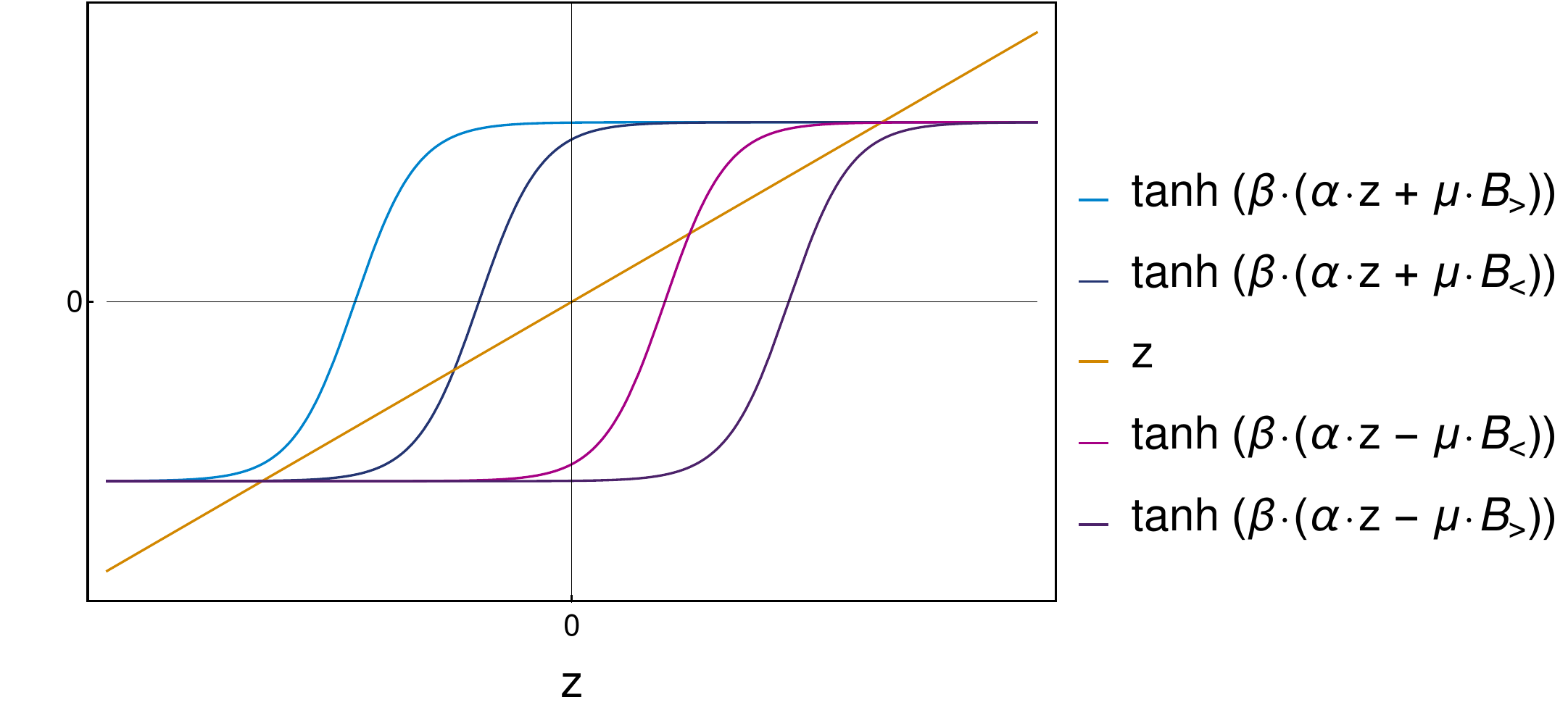} 
 \caption{Graphical illustration of the right-hand side of equation \eqref{zdotsaddlenode}, for $\alpha\,\beta \,>\,1$ and $\alpha\,\beta \,>\,1$ and $0\,<\,B^{<}\,<\,|B_\text{crit}|\,<\,B^{>}$
. When $|B| \,<\,|B_\text{crit}|$, there are three fixed point, two of them merge together (for $B\to\pm \,B_\text{crit}$) and vanish (for $|B|\,>\,|B_\text{crit}|$), hence obtaining a saddle node bifurcation. }
\label{SaddleNode_Graphics} 
\end{center}
\end{figure}

When the medium in which the spins are embedded is isotropic, there is no easy axis with respect to which we can choose the Lindblad operators. Nevertheless, the magnetic field field $\bB$ now imposes a preferred direction, which we can define to be the $z$ direction. If initially $\bm$ is not parallel to $\bB$, the effective field that a spin sees is the superposition of $\bB$ and the mean field $\bm$, and the Lindblad operators must be chosen according to the orientation of that effective field.  The dynamical equation of $\bm$ in this more general case will be treated further below in Section \ref{SubSection_2_dim_case}. However, as $\bm$ will become parallel to $\bB$ after a relaxation time, the dynamical equation \eqref{zdotsaddlenode} describes what happens in the subspace relevant for the bifurcation.

\section{Two-dimensional case} \label{SubSection_2_dim_case}

When in equation \eqref{eq_Lindblad_components} $z=0$ is an attracting value for the $z$ dynamics irrespective of the values of $x$ and $y$, we obtain an effectively two-dimensional system. A sufficient criterion for this to happen is that $h_{11}=h_{22}$ and $h_{23} = -h_{13}^*$ and $H=0$. Then the system of equations \eqref{eq_Lindblad_components} becomes (if we set $z=0$) 
\begin{eqnarray} \label{InitalValue_2dim}
\dot{\x}&=& 2 \, \text{Re}[h_{23} - h_{13}] + \left(\text{Re}[ h_{12}] - \Gamma \right) \, \x  -\text{Im} [h_{12}] \,\y . \\ \nonumber 
\dot{\y}&=& 2 \, \text{Im}[h_{23} + h_{13}]-\left( \text{Re}[ h_{12}] + \Gamma \right) \, \y  -\text{Im} [h_{12}]\,\x . 
\end{eqnarray}
    
From a physical point of view, a decaying value for $\z$ means that there is no equilibrium polarization along the $z$ axis. The probabilities for measuring spin-up in the $z$ direction becomes identical to that for measuring spin-down, namely 1/2. In the geometric view of the Bloch sphere, this means that the dynamics becomes restricted to a circle of radius one, namely the intersection of the Bloch sphere with the $z=0$ plane.

Now, there are two qualitatively different possible dynamical scenarios in the vicinity of a fixed point: If the Jacobian matrix corresponding to equations \eqref{InitalValue_2dim} has two real eigenvalues $\lambda_1 < \lambda_2 < 0$, we can call the eigenspaces corresponding to $\lambda_1$ and $\lambda_2$ the fast and slowly decaying directions. After some time, the dynamics along the fast direction has relaxed, and we have an effectively one-dimensional system along the slow direction. This is similar to the situation given in equations \eqref{eq_Lindblad_components_2} above. When a bifurcation occurs, the larger of the two eigenvalues goes through zero, which means that the bifurcation occurs within the one-dimensional subspace given by the slow direction, and all results obtained in the previous section can be applied also to this two-dimensional situation.
In the degenerate case where the two eigenvalues coincide (i.e., when the fixed point becomes a "star" \cite{strogatz2018nonlinear}), the dynamics follows a straight trajectory with the distance to the fixed point fully characterizing the dynamics, i.e., the dynamics is again effectively one-dimensional \cite{strogatz2018nonlinear}. 

The second scenario is the one where the two eigenvalues of the Jacobian matrix are complex conjugate. In this situation, the dynamical system can exhibit a Hopf bifurcation, where a stable spiral becomes unstable and a limit cycle (a closed, isolated trajectory) is generated. 
The normal form of the Hopf bifurcation is 
\begin{align} \label{InitalValue_Hopf3}
\dot{\x} =& \, \epsilon \, \x - \, b \, \y - \x \,r^2 \nonumber \\
\dot{\y} =& \, \epsilon \, \y + \, b \, \x - \y \,r^2\, ,
\end{align}
or, in polar coordinates,
\begin{align}\label{eq_2dim_Hopfradial}
\dot{r} =& \, \epsilon \,r -r^3 \\
\dot{\phi} =& \, b \, .\nonumber
\end{align}

If we impose the condition $|\epsilon| < 1$, the value of $r$ remains in the interval $[0,1]$. The bifurcation occurs at $\epsilon = 0$. There are no restrictions on the rate of angular increase $b$.

In the following we show that this normal form can be obtained by choosing suitable expressions for the $h_{ij}$ that ensure furthermore that the matrix $h_{ij}$ is positive semi-definite. For the diagonal elements, we set $\Gamma = a+r^2$ and $h_{11}=h_{22}= 2h_{33} = \frac \Gamma 2$. We will see that for sufficiently large $a$ the matrix $h_{ij}$ is positive semi-definite. For the nondiagonal elements, we write $\text{Re}[ h_{12}] =\kappa$ and set $\text{Im}[ h_{12}] = 0$. Together with  the choice 
$$\text{Re}[ h_{23}] =-\text{Re}[ h_{13}] = \frac{(\epsilon+a-\kappa)x - by}{4}$$ and 
$$\text{Im}[ h_{23}] = \text{Im}[ h_{13}] =\frac{(\epsilon+a+\kappa)y + bx}{4}$$ equations \eqref{InitalValue_2dim} take the normal form \eqref{InitalValue_Hopf3}.

The ansatz for the diagonal elements $h_{ii}$  ensures that the coefficient matrix $h$ is positive semi-definite for all $x,y \in [-1,1]$ if $a$ is large enough compared to $|\kappa|, |\epsilon|, |b|$.  
A sufficient criterion is that all principal minors of $h$ are non-negative \cite{horn2012matrix}. Hence, the following inequalities must be satisfied for all $x,y$: 
\begin{align} \label{PositiveSemiDefiniteMatrix}
h_{11} \geq& \,0 \nonumber \\
h_{22} \geq& \, 0 \nonumber  \\
h_{33} \geq& \,  0 \nonumber \\
h_{11} \cdot h_{22} - \vert h_{12} \vert^2 \geq&  \, 0 \nonumber \\ 
h_{11} \cdot h_{33} - \vert h_{13} \vert^2 \geq&  \, 0 \nonumber \\ 
h_{22} \cdot h_{33} - \vert h_{23} \vert^2 \geq&  \, 0 \nonumber \\ 
\text{Det} (h) = h_{11} \, h_{22} \, h_{33} + 2 \text{Re}[h_{12} \, h_{23} \, h_{31}] - h_{22} \,\vert h_{13} \vert^2 - h_{11} \,  \vert h_{23} \vert^2 -  h_{33} \,\vert h_{12} \vert^2\geq&  \, 0
\end{align}

By considering the leading terms in powers of $a$, it can be verified that these conditions are indeed fulfilled when $a$ is large enough. Here, the evaluation of $\text{Det} (h)$ is the most complicated one as not only the highest order in $a$, which is $a^3$, but also the next order, which is $a^2$ must be considered if the conditions shall be satisfied for all $r\in[0,1]$: We have
\begin{align*}
    \text{Det} (h) = \frac{a^3}{16}(1-r^2) + \frac{a^2r^2}{16}(3-r^2-2\epsilon) + \mathcal{O}(a).
\end{align*}
For $r=1$ the first term vanishes, and we must require that the second term is positive. This is satisfied for all $r\in[0,1]$ if $\epsilon<1$, for sufficiently large $a$. 
 the system one-dimensional. In the above example for a limit cycle, the rotating frame has an angular velocity of $b$. Within this rotating frame dynamics is one-dimensional with the relevant variable being $r$, which approaches the fixed point $r^* = \sqrt{\epsilon}$. 

Next, we construct again a microscopic model that undergoes such a bifurcation. To this purpose, we let the system of coupled spins interacts with an environment that produces a magnetic field in response to the magnetization of the spin system, with the field pointing in a direction that is rotated with respect to that of the magnetization. We make again the mean-field approximation.
The Hamiltonian for one spin is thus again (compare \eqref{eq_hmb})
\begin{equation} \label{h_hopf}
  \hat{h} = -\alpha \bm \cdot \bsigma - \mu \bB \cdot \bsigma \equiv -\mu \bB_{\text{eff}}(\bm) \cdot \bsigma\, ,
\end{equation}
where we have defined an effective field $\bB_{\text{eff}} = \bB  + \frac \alpha \mu \bm$.
Without loss of generality we choose the $x$-$y$ plane to be the plane spanned by the initial state of  $\bm$ and $\bB$.  
The strength of the field $\bB$  depends on $r$ and should vanish for $r=0$. We therefore set $|\bB| = \text{const} \cdot |\bm| + o(|\bm|)$.

Its direction is given by $ \be_{\phi+\delta_1}$, with $\phi$ being the direction in which $\bm$ points, and $\delta_1$ being the angle by which $\bB$ is rotated with respect to $\bm$. The effective field $\bB_{\text{eff}}$ is therefore also rotated with respect to $\bm$, but by a smaller angle that we call $\delta$. Such a sustained rotation of the magnetic field constitutes an active driving of the system, and it therefore requires an energy input into the device that measures $\bm$ and generates the field $\bB$. 

\begin{figure}[H]
\begin{center}
  \includegraphics[width=0.4\columnwidth]{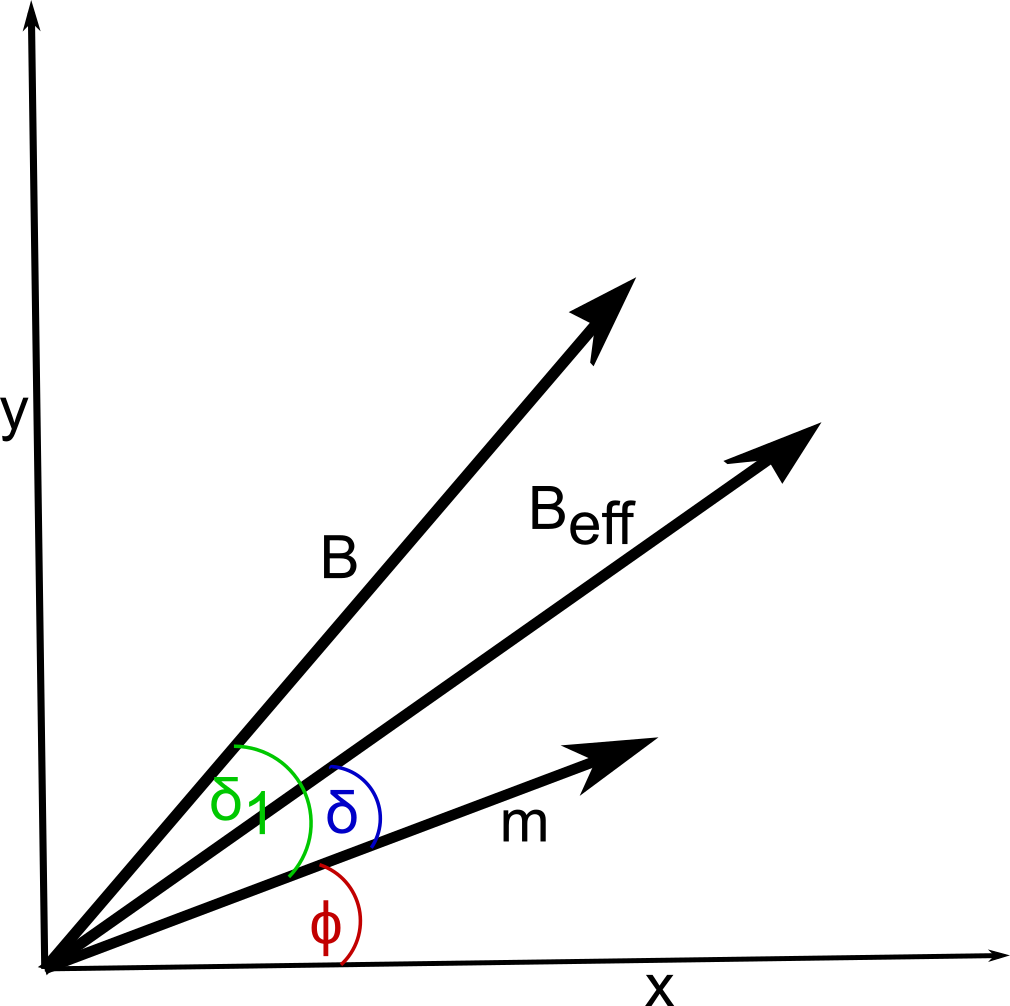} 
\caption{ Illustration of the angles $\phi$, $\delta$ and $\delta_1$ that describe the directions of the vectors $\bm$, $\bB$ and $\bB_{\text{eff}}$ }
\label{IllustrationOfAngles} 
\end{center}
\end{figure}

We assume again that the medium in which the spins are embedded has a finite temperature $T$, and that the interaction with the phonons of the medium causes spin flips between the two energy eigenstates of the spin. We denote the two rates by $\gamma_{f-}$ for the flip from the lower-energy state to the higher-energy state and $\gamma_{f+}$ for the reverse flip,
\begin{equation}\gamma_{f-}= \gamma_f\frac{e^{-\beta|\alpha\bm + \mu \bB|}}{e^{-\beta| \alpha\bm+\mu \bB|} + e^{\beta| \alpha\bm+\mu \bB|}} \quad \text{ and } \gamma_{f+} = \gamma_f\frac{e^{\beta| \alpha\bm+\mu \bB|}}{e^{-\beta| \alpha\bm+\mu \bB|} + e^{\beta| \alpha\bm+\mu \bB|}}\, .
\label{eq_fliprates}
\end{equation}
The sum of the two spin-flip rates is $\gamma_f$. 
In addition to the spin flips, the interaction with the phonons can also result in a measurement of the energy eigenstate, and we label the rate of this process with $\gamma_m$, where the letter $m$ stands for "measurement". 
The Hamiltonian (\eqref{h_hopf} does not only affect the flip rates, but it enters also the von Neumann term. In contrast to the previous examples, the density matrix is not diagonal in the eigenbasis of $\hat h$, which means that the von Neumann term does not vanish but causes a precession of the spin around the direction of $\bB$ . 
This makes the system three-dimensional. For the following analytical calculation, we therefore make the assumption that the rates $\gamma_m$ and $\gamma_f$ are so large that the unitary dynamics according to the von Neumann term can be neglected. In this case the vectors $\bm$ and $\bB$ stay in the $x$-$y$ plane. But we will additionally show below the result of a numerical integration of the full Lindblad equation with the von Neumann term in order to illustrate that the dynamics  still converges to a limit cycle, but with its center shifted vertically. 
The density matrix of the system has then the form
\begin{equation}
  \rho = \frac 1 2 
\begin{pmatrix}
1 & re^{-i\phi}\\
re^{i\phi} & 1
\end{pmatrix}
\end{equation}
with $r$ and $\phi$ being the representation of $\bm$ in polar coordinates in the $x$-$y$ plane.

In order to obtain the Lindblad equation that describes the dynamics of the density matrix $\rho$, we need to calculate the two operators that flip the spin between the + and - orientation with respect to an angular direction $\be_{\tilde\phi}$ in the $x$-$y$ plane, and the operator that performs a projection on an energy eigenstate. The latter one was not needed in the one-dimensional model as the density matrix was diagonal in the basis of the Hamiltonian. But when a spin is not in an eigenstate of $\hat h$, the interaction with the heat bath (phonons) will bring it into a state with definite energy, in addition to inducing flips between the two possible energy eigenstates, depending on whether an interaction event with the heat bath leads to the absorption or emission of a phonon, or not. The required rotation operator is given by (with $\bn$ being the unit vector parellel to the axis of rotation)
\begin{equation}
\textbf{R} = \e^{-i{\tilde\phi} {\bn}\cdot\bsigma/2} = \e^{-i{\tilde\phi} \sigma_z /2} = 
\begin{pmatrix}
\e^{-i{\tilde\phi}/2} & 0\\
0 & \e^{i{\tilde\phi}/2} 
\end{pmatrix} \, .
\end{equation}
For a spin that flips between the $+x$ and $-x$ direction, the flip matrices are
\begin{equation}
\textbf{F}_x = \begin{pmatrix}
  \frac 1 2 & -\frac 1 2\\
  \frac 1 2 & -\frac 1 2
\end{pmatrix}
\qquad \text{ and } \qquad \textbf{F}_x^\dagger = \begin{pmatrix}
  \frac 1 2 & \frac 1 2\\
  -\frac 1 2 & -\frac 1 2
\end{pmatrix} \, .
  \end{equation}
  The measurement matrix is 
  \begin{equation}
\textbf{M}_x = \begin{pmatrix}
  0 & 1\\
   1  & 0
\end{pmatrix} \, .
\end{equation}
These are the matrices with respect to the direction ${\tilde\phi} = 0$ in the $x$-$y$ plane. Performing a rotation by an angle ${\tilde\phi}$ 
gives the general measurement matrix with respect to the direction of $\bB_{\text{eff}}$,
\pagebreak[1]
\begin{equation}
\textbf{M} =\textbf{R}\textbf{M}_x \textbf{R}^\dagger = \begin{pmatrix}
   0 & e^{-i\tilde\phi}\\
  e^{i\tilde\phi} & 0
\end{pmatrix}
\end{equation}
and the flip matrix
\begin{equation}
  \textbf{F} = \textbf{R} \textbf{F}_x \textbf{R}^\dagger =
  \frac 1 2 
  \begin{pmatrix}
1 & -e^{-i{\tilde\phi}}\\
e^{i{\tilde\phi}} & -1
\end{pmatrix}
\end{equation}
and the associated matrix $\textbf{F}^\dagger$.

The Lindblad equation with the three Lindblad operators $\textbf{M}$,  $\textbf{F}$ and $\textbf{F}^\dagger$ is

\begin{eqnarray}\label{Lindblad_explizit_Hopf}
 \dot \rho &=& \gamma_m\textbf{M} \rho \textbf{M}^\dagger + \gamma_{f+} \textbf{F} \rho \textbf{F}^\dagger + \gamma_{f-} \textbf{F}^\dagger \rho \textbf{F}
 -\frac{\gamma_{m}} 2 \left\{\textbf{M}^\dagger \textbf{M},\rho\right\}   -\frac{\gamma_{f+}} 2 \left\{\textbf{F}^\dagger \textbf{F},\rho\right\}   -\frac{\gamma_{f-}} 2 \left\{\textbf{F} \textbf{F}^\dagger,\rho\right\}\nonumber \\
  &=& \frac {\gamma_{f+}}{4}   \begin{pmatrix}
0 & X\\
X^* & 0
\end{pmatrix} + \frac {\gamma_{f-}}{4}   \begin{pmatrix}
0 & Y\\
Y^* & 0
\end{pmatrix} 
+ \frac{\gamma_m} 2  \begin{pmatrix}
0 & Z\\
Z^* & 0
\end{pmatrix} 
\end{eqnarray}  
with 

\begin{align*}
X =& \;\; \;\;\; 2\, \e^{-i{\tilde\phi}}- r \left((\cos(\phi - {\tilde\phi}) e^{-i{\tilde\phi}} + e^{-i\phi}\right) \\
Y =& \,-2\, \e^{-i{\tilde\phi}} - r \left(\cos(\phi - {\tilde\phi}) e^{-i{\tilde\phi}} + e^{-i\phi}\right) \\
Z =& \, r \, \e^{-i\phi} \, \left(e^{-2i(\tilde \phi - \phi)}-1\right)
\end{align*}

When we set ${\tilde\phi} = \phi + \delta$ and consider the nonzero matrix elements of the Lindblad equation, we obtain
\begin{equation}
\frac d{dt}(re^{-i\phi}) =  \gamma_m re^{-i{\phi}}(e^{-2i\delta}-1) + {e^{-i(\phi + \delta)}}(\gamma_{f+} - \gamma_{f-}) - \frac{r}{2}(\gamma_{f+} + \gamma_{f-})e^{-i\phi}(1+e^{-i\delta}\cos\delta)\, .
\end{equation}
Separating the real and imaginary part on the left- and right-hand side and setting $\gamma_{f+} - \gamma_{f-} = \Delta \gamma_f$ and $\gamma_{f+} + \gamma_{f-} = \gamma_f$, this gives the two equations

\begin{eqnarray}\label{eq_hopf_example}
  \dot r &=& {\Delta \gamma_f} \cos \delta - \frac r 2 \gamma_f(1+\cos^2\delta) -2r\gamma_m \sin^2\delta \nonumber\\
  \dot \phi &=&  \frac {\Delta \gamma_f}{r} \sin\delta 
    - \frac{\gamma_f}{2} \cos \delta \sin \delta + \gamma_m \sin(2\delta)\, .
\end{eqnarray}
For $\delta = 0$, the radial equation reduces to \eqref{dotzpitchfork} for the one-dimensional system, as then  $r=z$. Since $r$ is positive, we obtain only the positive branch of \eqref{dotzpitchfork}. 

Since we have now assumed that $\bB$ depends on $\bm$ and that it vanishes when $\bm$ vanishes, $\Delta \gamma_f$ must be proportional to $r$ to leading order in $r$  ($\Delta \g_f = \g_{f+}-\g_{f+} = \g_f \, \tanh\left[ \beta |\alpha \, \bm + \mu \, \bB | \right] = \g_f \, \tanh\left( \beta \, c \, r \right) = \g_f \, \beta \, c \, r + \mathcal{O}(r^3)$ ), and $r=0$ is always a fixed point. 

 If the fixed point at $r=0$ is unstable, we have a stable fixed point $r^*\in(0,1)$, which is determined by the equation
\begin{equation}
  {\Delta \gamma_f} \cos\delta = \frac{r^*}{2} \gamma_f(1+\cos^2\delta) + 2\, r^* \gamma_m \sin^2\delta\, .
  \end{equation}
In this case we obtain
\begin{equation}
  \dot \phi = \frac {\gamma_f}{ 2} \tan \delta + 2\,\gamma_m \tan \delta\, .
\end{equation}
The angle $\phi$ increases at a constant rate.  For the temperature $T$ (or, equivalently, the value of $\beta$) at which the stable stationary solution for $r$ becomes nonzero, a Hopf bifurcation occurs. 
As a side result, we have shown that the change of the direction of $\bm$ is towards $\bB$. If the magnetic field was constant in time, $\bm$ would relax towards it. This completes the remark made at the end of section \ref{SubSection_1_dim_case}.

\begin{figure}
    \centering
\includegraphics[width = 0.6 \textwidth]{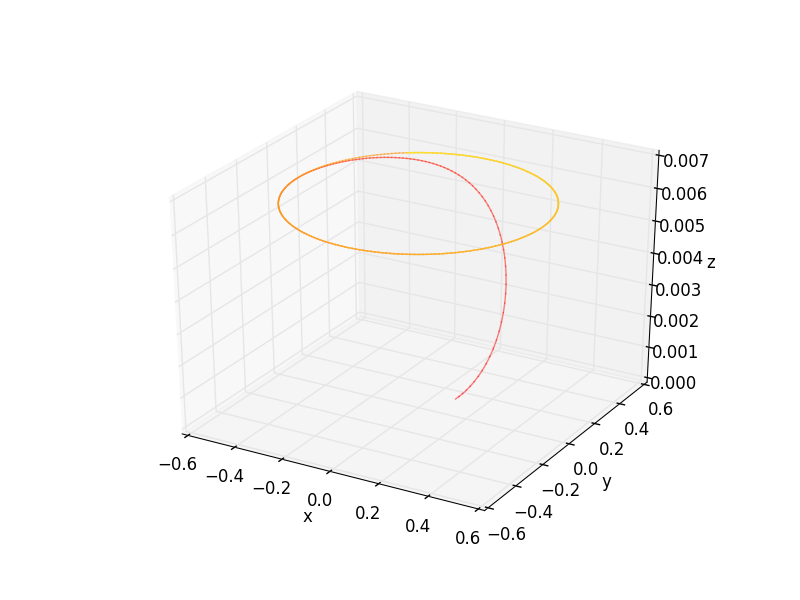}
    \caption{An example trajectory obtained with the Hamiltonian \eqref{h_hopf}, where the magnetic field $\bB(\bm)$ points in the direction that is obtained by rotating $\bm$ by an angle $\delta_1$ around the $z$ axis, and its size is proportional to $|\bm|$. The equations of motion \eqref{eq_hopf_example} are supplemented by the von Neumann term and generalized to three dimensions, as explained in Section \ref{SubSection_3_dim_case}. The parameter values are $\gamma_f = 2.8$, $\gamma_m=2$, $\delta_1 = 0.3$, $(\mu|\bB|)/(\alpha|\bm|) = 0.1$, $\beta\alpha=1$, $\hbar/\alpha = 10$. As time proceeds, the color changes from red to yellow, and the trajectory converges to a limit cycle.}
    \label{fig_hopf}
\end{figure}
Figure \ref{fig_hopf} shows the trajectory obtained by a numerical integration of the model for a value of $\beta$ large enough to obtain a limit cycle. When performing this simulation, we additionally included the von Neumann term in the Lindblad equation. We see a limit cycle that is shifted in the $z$ direction. Without the von Neumann term, the limit cycle would be exactly in the $x$-$y$ plane. 

\section{Three-dimensional case} \label{SubSection_3_dim_case}

We now look at the full system of equations \eqref{eq_Lindblad_components} with no constraint forcing the dynamics onto a lower-dimensional manifold. A nonlinear three-dimensional dynamical system can exhibit the same types of bifurcations as the lower-dimensional systems, but in addition it can make the transition to chaos and show a strange attractor \cite{strogatz2018nonlinear}. 
In general, we can expect that   equations of the type \eqref{eq_Lindblad_components} display also strange attractors. In order to demonstrate this explicitly, we construct a specific model that is similar in spirit to the previous ones. We use again a system of interacting spins in mean-field approximation coupled to a heat bath and subjected to a magnetic field the direction and strength of which depends on the magnetization $\bm$. As in the previous section, the interaction with the phonons causes the spins to make transitions to and between their energy eigenstates. 

The main idea now is to choose the function $\bB(\bm)$ such that it makes $\bm$ move along a trajectory that bears similarity to that of a classical strange attractor.Similarly to the  previous example, such an ongoing change of $\bB$ in response to $\bm$ leads to the dissipation of energy in the system and can be sustained only by providing an energy supply. Since $|\bm|$ is confined to the interval $[0,1]$, it is important to choose parameters such that $|\bm|$ is not close to 1 at all times since it will then hardly respond to a change of the strength $B$ of the magnetic field and can then not show a chaotic trajectory. We choose the cyclically symmetric Thomas model \cite{thomas1999deterministic}
as inspirational source and set 
\begin{eqnarray}\label{thomas}
\frac \mu\alpha B_x &=& a\sin(m_y/A) \nonumber\\
\frac \mu\alpha B_y &=& a\sin(m_z/A) \\
\frac \mu\alpha B_z &=& a\sin(m_x/A)\, .\nonumber
\end{eqnarray}

At any moment in time, we can define a two-dimensional plane spanned by the vectors $\bm$ and $\bB$. During an infinitesimal time interval $dt$ the dissipative dynamics according to the flip and measurement matrices cause a change $d\bm$ of $\bm$ towards $\bB$, with the angular change of $\bm$ in this plane and the radial change being given by the equations \eqref{eq_hopf_example}. Since the plane within which this motion takes place changes with time, the equations of motion \eqref{eq_hopf_example} must now be transformed to a form that is independent of the coordinate system. We have $r = \sqrt{x^2+y^2+z^2} = |\bm|$, and the direction of angular change is that of the vector $(\bm \times \bB) \times \bm$. We therefore write \eqref{eq_hopf_example} in the form
\begin{equation}\label{eq_m_allg}
 \dot {\br} =   \dot {\bm} = \frac{\bm(t)}{ m} \dot{r} + \frac {\left(\bm \times \bB\right) \times \bm }{|\left(\bm \times \bB\right) \times \bm|} \, m \, \dot \phi \equiv \be_r(t) \dot r + m \be_\phi(t) \dot \phi\, .
\end{equation}
Here, $\phi$ is no longer an angle in the $x$-$y$ plane, but in the plane spanned by $\bm$ and $\bB$. With this definition of $\phi$, the expressions for $\dot r$ and $\dot \phi$ are given by \eqref{eq_hopf_example} and the flip rates by \eqref{eq_fliprates}, as before.

We also want to consider the case that the von Neumann term cannot be neglected. We have already given a component-wise version of the von Neumann term as the last terms in equations \eqref{eq_Lindblad_components}. Evaluating these terms with the Hamiltonian \eqref{h_hopf} gives the final version of Lindblad equation (remember that $|\bm| = r = \sqrt{x^2+y^2+z^2}$ and $\bB_{\text{eff}} = \bB  + \frac \alpha \mu \bm$)
\begin{eqnarray}\label{eq_lindblad_chaos_full}
  \dot x &=&  (\be_r)_x \, \dot r + (\be_\phi)_x\,r  \dot \phi + \frac {2\mu}{\hbar}\left(y\,B_{\text{eff}}^{(z)}-z\,B_{\text{eff}}^{(y)}\right)\nonumber \\
  \dot y &=& (\be_r)_y \, \dot r + (\be_\phi)_y\,r  \dot \phi + \frac {2\mu}{\hbar} \left(z\,B_{\text{eff}}^{(x)}-x\,B_{\text{eff}}^{(z)}\right) \\
  \dot z &=&  (\be_r)_z \, \dot r + (\be_\phi)_z\,r  \dot \phi + \frac {2\mu}{\hbar}\left(x\,B_{\text{eff}}^{(y)}-y\,B_{\text{eff}}^{(x)}\right)\nonumber 
  \end{eqnarray}

This dynamics of the density matrix with the choice \eqref{thomas} for $\bB(\bm)$ was evaluated by numerical integration. 

\begin{figure}[H]
\begin{center}
  \includegraphics[width=0.45\textwidth]{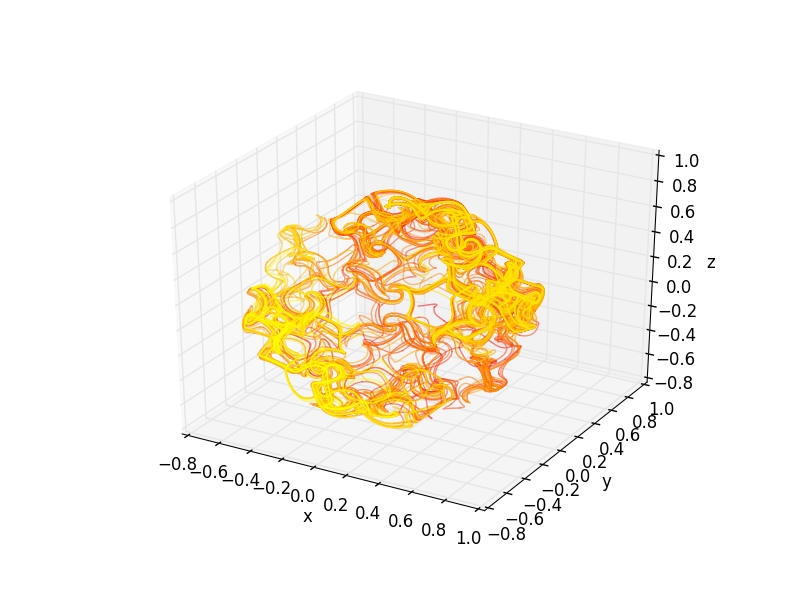} 
  \includegraphics[width=0.45\textwidth]{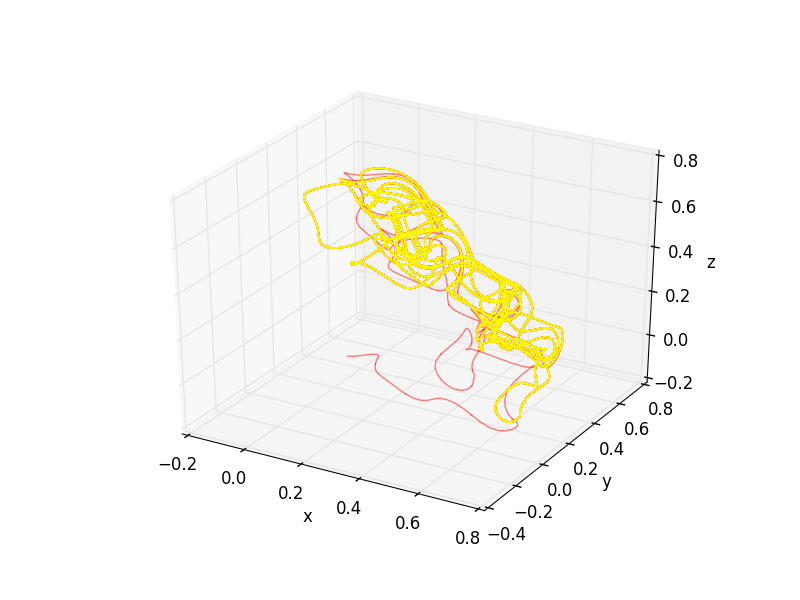} 
 \caption{The dynamics of model \eqref{eq_lindblad_chaos_full} with the magnetic field chosen according to \eqref{thomas}, with the parameters $a=0.1$, $A=0.05$, $\gamma_f=2.8$, $\gamma_m =2$, $\beta\alpha =1.3$, $\hbar/\alpha = 200$ (left) and $\hbar/\alpha = 20$ (right). As time progresses, the color changes continuously from red to yellow. Increasing the von Neumann term eventually destroys the chaotic attractor and leads to a limit cycle for this parameter choice.}
 \label{Fig_Chaos}
\end{center}
\end{figure}

Figure \ref{Fig_Chaos} shows the trajectory of $\bm$ obtained with a parameter set that leads to chaos when the von Neumann term is sufficiently small compared to the dissipative terms in the left graph, and a limit cycle when the von Neumann term is larger.

\section{Discussion and Conclusion}

We have studied the dynamics of Lindblad equations where the transition rates are nonlinear in the density matrix and have constructed explicit models that show such a dynamics. This approach raises the interesting issue of how the density matrix should be interpreted. In the conventional ensemble interpretation, the Lindblad equation must be linear in the density matrix. Often, the density matrix is interpreted as a device to reflect our limited knowledge of the precise state a quantum system, which is assumed to be a pure state. This applies in particular to the ensemble description of statistical mechanics systems, where the probabilistic description is often viewed as consequence of our ignorance of the exact quantum state of the system. In this case, the time evolution must be linear in the density matrix since the different pure states of the ensemble cannot interact with one another.

Decoherence theory, which is often taken as the background theory for deriving the Lindblad equation, is based on a similar assumption: here, the system is considered as part of a larger system that includes an environment with which it interacts, and the dynamics of  this larger system is taken to be unitary. Due to the interaction, the system becomes entangled with the bath. Assuming a sufficient amount of randomness or uncorrelatedness of the environmental degrees of freedom, the environmental state that co-occur with the difference system states, can be argued to be orthogonal to each other. This is the central assumption of decoherence theory. The full quantum state of system (S) and environment (E) thus has the form $\sum_n c_n\ket{\psi}^{(\text{S})}_n \ket{\Psi}^{(\text{E})}_n$ with orthogonal $\{\ket{\Psi}_n^{(\text{E})}\}$. The reduced density matrix of the system, which is obtained from the full density matrix by taking the trace over the environmental states, is then that of the ensemble of the  $\ket{\psi}^{(\text{S})}_n$, and therefore the time evolution of this reduced density matrix  is linear. This type of consideration is also applied in approaches to the quantum measurement problem \cite{Strunz2002dekoh/"arenz} and in the field of quantum foundations of statistical mechanics \cite{eisert2015quantum}. 

Although decoherence theory is by some authors considered as an explanation of the quantum-classical transition and in particular of the measurement problem \cite{Strunz2002dekoh/"arenz}, others emphasize that it cannot explain why only one of the different possible outcomes is realized in a one-time run of the measurement experiment \cite{adler2003decoherence,schlosshauer2007decoherence}, since the full state of system and environment is assumed to be an entangled state that contains all the $\ket{\psi}^{(\text{S})}_n$ at the same time. Furthermore, it is argued that a unitary time evolution of macroscopic systems is incompatible with the actual calculations performed in statistical mechanics and condensed matter theory \cite{drossel2020condensed}, and that the quantum-classical transition indicates that there are limits of validity to unitary time evolution \cite{leggett}.     

It is therefore apparent that the assumptions that lead to a time evolution that is linear in the density matrix are not applicable in all situations. In particular the idea that the system, possibly taken together with its environment, is in a pure quantum state, can be challenged, since in
most experimental situations, the preparation of a quantum mechanical state cannot be controlled in all detail. One reason for this is the nonzero temperature of the preparation device. In this situation, it is impossible even in principle to prepare a pure state that can be verified experimentally, for instance by quantum state tomography \cite{d2001quantum}. When considering this imprecision of a quantum state as fundamental, the density matrix does not only reflect our limited knowledge but is the best description of the system we can possibly give. This also means that the density matrix now can be understood as describing a single system and not an ensemble of systems. Interpreting the density matrix as describing a single system is also advocated by other authors \cite{anandan1999meaning}.

Furthermore, when not just the preparation, but also the subsequent time evolution of the quantum system is subject to uncontrollable stochastic influences, this density matrix does not evolve according to the von Neumann equation but requires a Lindblad equation (or a non-Markovian equation, but we do not consider this situation here). Since now the density matrix represents a single system and not an ensemble, the Lindblad equation is no longer required to be linear in the density matrix, and nonlinearities can arise under suitable circumstances.

In this paper, we have obtained nonlinearities in the Lindblad equation by applying a mean-field approximation to the interactions between spins. When performing a mean-field approximation on a quantum system, one usually replaces quantum mechanical operators by their expectation value, with the result that the influence of the other quantum particles take the shape of an external potential (which is a feature of classical physics). The nonlinearities arose in our model because this mean-field approximation made the one-particle Hamiltonian dependent on the density matrix, and this Hamiltonian in turn enters the von Neumann term as well as the temperature-dependent transition rates. 
We think that the reason why mean.-field like approximations are so widespread and successful is that they capture features of reality, which is quantum only up to certain (temperature-dependent) length scales. In our calculations, this approximation included furthermore the neglect of spatial correlations, making the one-dimensional versions of our model formally equivalent to the mean-field theory of ferromagnetism: 
According to the Landau theory of phase transitions, the free energy of a uniaxial ferromagnet in the presence of a magnetic field $h$ is
$F = \,r \, m^2 + u \, m^4-hm$, with $m$ being the order parameter (magnetisation) and with the parameters satisfying $u>0$ and $r \propto (T-T_C)$. Relaxation dynamics toward equilibrium takes the form
$$
\frac {dm}{dt} = -\lambda \frac {dF}{dm} = -2\lambda r m - 4 \lambda u m^3 -\lambda h\, ,
$$
which corresponds to \eqref{dotzpitchfork} for $h=0$ and to \eqref{zdotsaddlenode} for $h \neq 0$. 
Mean-field theories are a useful tool to capture by an analytical calculations the phenomenology of the overall behavior.  Since spatial correlations are neglected, they have their limitations, as they cannot describe phenomena such as the formation of domains, of spatiotemporal patterns or spatiotemporal chaos, or how fluctuations modify the dynamics.   
In addition to the pitchfork and saddle-node bifuractions, one-dimensional systems also show a transcritical bifurcation, where two fixed points exchange their stability, as one of them moves into the physically feasible coordinate region. An example for a simple model that shows this bifurcation is the Laser model by Haken (see for instance \cite{strogatz2018nonlinear}). As it is different in nature from the finite-temperature mean-field spin models on which we focused here, we did not include the transcritical bifurcation in Section \ref{SubSection_1_dim_case}.  

In order to obtain limit cycles and chaos, we had to introduce a feedback between the magnetic field $\bB$  and the  entries $(x,y,z)$ of the density matrix.  If a system of coupled spins in a magnetic field is left to itself and its interaction with a heat bath, an equilibrium state with detailed balance must be reached, and this is not compatible with a sustained change of the magnetization of the coupled spin system. Sustained changes require an ongoing driving of the system away from equilibrium, and this implies that the feedback mechanism requires an ongoing energy supply. This is a comparable situation to the quantum state transformation experiments described in the Introduction. There, the apparatus that prepares and measures or selects the quantum states is also a classical device that is connected to a power supply.

One can also think of other ways to drive the quantum system to a periodic of chaotic trajectory. For example, one could apply fields that do not change in time, and add an active feedback process that prevents relaxation away from an unstable trajectory. Such a stabilization can be done in a minimally invasive way \cite{pyragas2002stabilizing}. Another and somewhat trivial way to drive a quantum system on a periodic or chaotic trajectory would be to impose this trajectory by top-down control. If we place a quantum two-level system in an environment that has a magnetic field and a temperature that changes on a time scale that is much slower than the thermal equilibration of the quantum system, we obtain, 
\begin{align} \label{vecxBT}
\br(t) = \frac{\bB(t)}{|\bB(t)|} \tanh{\left(\beta \mu |\bB(t)| \right)} \, .
\end{align}
 In such a situation, the density matrix is well defined at every moment in time, as is also argued by authors working in the field of NMR \cite{paniagua2006physical}. A system that leads to a chaotic equation of motion for the density matrix of the spin can be built in the following way: First, one constructs a electromechanical device that leads to a chaotic equation of motion in spherical coordinates $r$, $\theta$, $\phi$ of the end point of an arm of variable length and orientation. Then one mounts a magnet on the end of this arm such that the field points inwards in radial direction, and one places the quantum spin system in the center at $r=0$. The direction and strength of the magnetic field at the location of the spin are then fixed by the angle $(\theta,\phi)$ and radius $r$ respectively. This means that the equation of motion of the electromechanical device translates into an equation of motion of the magnetic field at the center, and this in turn translates into an equation of motion of the type \eqref{eq_Lindblad_components} for the density matrix of the spin. 

The last example has demonstrated most clearly that the density matrix is determined by the environment of the quantum system. In this example, the environment is given by a classical device that determines the magnetic field and the temperature to which the quantum system is exposed. 
There is no other way in which a quantum system can be controlled or influenced apart from  changing classical, macroscopic control parameters, which then can cascade down to the quantum system.  
When the environment is simple, for instance a heat bath that does not change in time, this top-down determination of the density matrix is often overlooked, although temperature is a thermodynamic variable that is imposed by the environment. In the examples discussed in Sections \ref{SubSection_2_dim_case} and \ref{SubSection_3_dim_case}, there was additionally a bottom-up influence from the quantum system to the classical environment, as the classical environment responded to the magnetization of the system. In quantum-state transformation protocols, there is also such a feedback from the quantum system to the classical system, as the postselection depends on the measurement outcome. The difference to our examples is that in our mean-field model the measurement is done on all interacting spins of the system, which leads to a classical measurement result (the vector $\bm$), and not to the binary outcome of the measurement on a spin-1/2 system.

To conclude, the interaction of a quantum system with a classical system can lead in various ways to a violation of the unitary dynamics of wave functions as well as to the linear dynamics of the density matrix, both of which are only valid under restricted conditions and assumptions. Since our world is full of nonlinear dynamical phenomena, linear dynamics can only be a special case, even though a lot of research focuses on this case.  
It is our belief and that of various other authors \cite{ellis2012,drossel2018contextual,grangier2018quantum} that the top-down influence from the classical world on the quantum world is an irreducible feature of nature that must be taken into account when one wants to find a solution to the puzzles surrounding the interpretation of quantum mechanics.

\bibliographystyle{ieeetr}
\bibliography{quantumlit} 

\begin{thebibliography}{10}

\bibitem{Ballentine}
L.~E. Ballentine, ``The statistical interpretation of quantum mechanics,'' {\em
  Rev. Mod. Phys.}, vol.~42, pp.~358--381, Oct 1970.

\bibitem{alicki2007quantum}
R.~Alicki and K.~Lendi, {\em Quantum dynamical semigroups and applications},
  vol.~717.
\newblock Springer, 2007.

\bibitem{rivas2012open}
A.~Rivas and S.~F. Huelga, {\em Open quantum systems}.
\newblock Springer, 2012.

\bibitem{lindblad1976generators}
G.~Lindblad, ``On the generators of quantum dynamical semigroups,'' {\em
  Communications in Mathematical Physics}, vol.~48, no.~2, pp.~119--130, 1976.

\bibitem{maassen2004quantum}
H.~Maassen, ``Quantum probability quantum information theory quantum
  computing,'' {\em Lecture notes of a course to be given in the spring
  semester of}, 2004.

\bibitem{drossel2020condensed}
B.~Drossel, ``What condensed matter physics and statistical physics teach us
  about the limits of unitary time evolution,'' {\em Quantum Studies:
  Mathematics and Foundations}, vol.~7, pp.~217--231, 2020.

\bibitem{leggett}
A.~J. Leggett, ``Testing the limits of quantum mechanics: motivation, state of
  play, prospects,'' {\em Journal of Physics: Condensed Matter}, vol.~14,
  no.~15, p.~R415, 2002.

\bibitem{breuer2002theory}
H.-P. Breuer and F.~Petruccione, {\em {\it The theory of open quantum
  systems}}.
\newblock Oxford University Press on Demand, 2002.

\bibitem{bechmann1998non}
H.~Bechmann-Pasquinucci, B.~Huttner, and N.~Gisin, ``Non-linear quantum state
  transformation of spin-12,'' {\em Physics Letters A}, vol.~242, no.~4-5,
  pp.~198--204, 1998.

\bibitem{kiss2006complex}
T.~Kiss, I.~Jex, G.~Alber, and S.~Vym{\v{e}}tal, ``Complex chaos in the
  conditional dynamics of qubits,'' {\em Physical Review A}, vol.~74, no.~4,
  p.~040301, 2006.

\bibitem{kiss2011measurement}
T.~Kiss, S.~Vym{\v{e}}tal, L.~T{\'o}th, A.~G{\'a}bris, I.~Jex, and G.~Alber,
  ``Measurement-induced chaos with entangled states,'' {\em Physical review
  letters}, vol.~107, no.~10, p.~100501, 2011.

\bibitem{torres2017measurement}
J.~M. Torres, J.~Z. Bern{\'a}d, G.~Alber, O.~K{\'a}lm{\'a}n, and T.~Kiss,
  ``Measurement-induced chaos and quantum state discrimination in an iterated
  tavis-cummings scheme,'' {\em Physical Review A}, vol.~95, no.~2, p.~023828,
  2017.

\bibitem{hartmann2017asymptotic}
M.~Hartmann, D.~Poletti, M.~Ivanchenko, S.~Denisov, and P.~H{\"a}nggi,
  ``Asymptotic floquet states of open quantum systems: the role of
  interaction,'' {\em New Journal of Physics}, vol.~19, no.~8, p.~083011, 2017.

\bibitem{strogatz2018nonlinear}
S.~H. Strogatz, {\em Nonlinear Dynamics and Chaos with Student Solutions
  Manual: With Applications to Physics, Biology, Chemistry, and Engineering}.
\newblock CRC Press, 2018.

\bibitem{horn2012matrix}
R.~A. Horn and C.~R. Johnson, {\em Matrix analysis}.
\newblock Cambridge university press, 2012.

\bibitem{thomas1999deterministic}
R.~Thomas, ``Deterministic chaos seen in terms of feedback circuits: Analysis,
  synthesis," labyrinth chaos",'' {\em International Journal of Bifurcation and
  Chaos}, vol.~9, no.~10, pp.~1889--1905, 1999.

\bibitem{Strunz2002dekoh/"arenz}
W.~Strunz, G.~Alber, and F.~Haake, ``Dekohärenz in offenen {Q}uantensystemen:
  {V}on den {G}rundlagen der {Q}uantenmechanik zur {Q}uantentechnologie,'' {\em
  Physik Journal}, vol.~1, no.~11, pp.~47--52, 2002.

\bibitem{eisert2015quantum}
M.~F. Jens~Eisert and C.~Gogolin, ``Quantum many-body systems out of
  equilibrium,'' {\em Nature Physics}, vol.~11, no.~2, pp.~124--130, 2015.

\bibitem{adler2003decoherence}
S.~L. Adler, ``Why decoherence has not solved the measurement problem: a
  response to {P. W. Anderson},'' {\em Studies in History and Philosophy of
  Science Part B: Studies in History and Philosophy of Modern Physics},
  vol.~34, no.~1, pp.~135--142, 2003.

\bibitem{schlosshauer2007decoherence}
M.~A. Schlosshauer, {\em Decoherence: and the quantum-to-classical transition}.
\newblock Springer Science \& Business Media, 2007.

\bibitem{d2001quantum}
G.~D'Ariano and P.~L. Presti, ``Quantum tomography for measuring experimentally
  the matrix elements of an arbitrary quantum operation,'' {\em Physical review
  letters}, vol.~86, no.~19, p.~4195, 2001.

\bibitem{anandan1999meaning}
J.~Anandan and Y.~Aharonov, ``Meaning of the density matrix,'' {\em Foundations
  of Physics Letters}, vol.~12, no.~6, pp.~571--578, 1999.

\bibitem{pyragas2002stabilizing}
K.~Pyragas, V.~Pyragas, I.~Kiss, and J.~Hudson, ``Stabilizing and tracking
  unknown steady states of dynamical systems,'' {\em Physical review letters},
  vol.~89, no.~24, p.~244103, 2002.

\bibitem{paniagua2006physical}
J.~C. Paniagua, ``On the physical interpretation of density operators at the
  atomic scale: A thorough analysis of some simple cases,'' {\em Concepts in
  Magnetic Resonance Part A: An Educational Journal}, vol.~28, no.~6,
  pp.~384--409, 2006.

\bibitem{ellis2012}
G.~F. Ellis, ``On the limits of quantum theory: Contextuality and the
  quantum--classical cut,'' {\em Annals of Physics}, vol.~327, no.~7,
  pp.~1890--1932, 2012.

\bibitem{drossel2018contextual}
B.~Drossel and G.~Ellis, ``Contextual wavefunction collapse: An integrated
  theory of quantum measurement,'' {\em New Journal of Physics}, vol.~20,
  no.~11, p.~113025, 2018.

\bibitem{grangier2018quantum}
P.~Grangier and A.~Auff{\`e}ves, ``What is quantum in quantum randomness?,''
  {\em Philosophical Transactions of the Royal Society A: Mathematical,
  Physical and Engineering Sciences}, vol.~376, no.~2123, p.~20170322, 2018.

\end{thebibliography}

\end{document}